\documentstyle[12pt]{article}
\begin{document}
\vskip -7em\hfill{\small SWAT/94-95/49 \break\vskip3em}
\begin{center}
{\Large \bf The Microscopic Representation of}
\end{center}
\begin{center}
{\Large \bf Complex Macroscopic Phenomena}
\end{center}
\begin{center}
{\bf Critical Slowing Down - A Blessing in Disguise}
\begin{center}
{(to appear in {\it Annual Reviews of Computational Physics II})}
\end{center}
\end{center}
\begin{center}
{\bf S. Solomon}
\end{center}
\begin{center}
{e-mail: sorin@vms.huji.ac.il}
\end{center}
\begin{center}
{\it Department of Physics, University College of Swansea}
\end{center}
\begin{center}
{and}
\end{center}
\begin{center}
{\it  Racah Institute of Physics,
Hebrew University of Jerusalem}
\end{center}
\begin{center}
{(permanent address)}
\end{center}
\begin{abstract}

Many complex systems are representable as macroscopic sets
of elements which interact by simple rules.

The complex macroscopically relevant phenomena are then the result
of the generic emergence of a space-time multi-scale dynamics.

Critical Slowing Down labels the emerging global features and
describes their complex collective evolution.

This paradigm is quite universal and extends
to a very wide range of systems and disciplines.

\end{abstract}

\section{The Microscopic Representation Paradigm}

If one throws a stone and  wishes to predict its trajectory
one needs only to consider it as a single rigid body.

If one wishes to study its light
spectrum, one has to consider only its atomic and molecular
structure.

If one wishes to study how the stone deforms under pressure,
one may treat it as an elastic continuum.

However, if one increases the pressure
and one wishes to know how it breaks,
one has to consider it as a conglomerate of
smaller stones (and stones within stones)
with cracks and faults developing over many
space and time scales down to the microscopic one.

The present paper concerns complex systems for which
the origin of complexity can be traced to the very attempt
by our perception
to describe a macroscopic number of microscopic
objects and events in terms of a limited number of
macroscopic features.

We will discuss the techniques through which one can
systematically follow the birth of the complex macroscopic
phenomenology out of the simple fundamental microscopic laws.

In the field of fundamental physics,
such understanding was obtained for a wide range
of phenomena using theories based on a
{ {\bf Microscopic Representation}} (in short,
in the following text {\bf MicRep}) paradigm.

The MicRep paradigm consists in deducing the
macroscopic objects ( {\bf Macro}s from now on)
and their phenomenological complex ad-hoc
laws in terms of a multitude of elementary
microscopic objects ({\bf Micro}s from now on) interacting
by simple fundamental laws.
The Macros and their laws emerge then naturally from the
collective dynamics of the Micros
as its effective global large scale features.

However, the mere microscopic representation of a system
cannot lead to a satisfactory and complete
understanding of the macroscopic phenomena.
Indeed, the mere copying on the computer
of a real-life system with all its problems
does not by itself constitute a solution to those
problems.

It is clear that a satisfactory MicRep procedure
of such complex systems has to be "multi-scale"
i.e.:
\begin{itemize}
\item one has to recognize the relevant objects
which describe effectively the system at each scale and
\item one has to establish the relations between the
objects and the laws prevailing at different scales.
\end{itemize}

Therefore, the MicRep approach is not trying
to substitute the study of
one scale for the study of another scale;
one is trying
to unify into a coherent picture the complementary
descriptions of a one and the same reality.

In fact one can have a multitude of scales
such that the Macros of one scale become the Micros of
the next level. As such the "elementary" Micros
of one MicRep do not need to be elementary in
the fundamental sense: it is enough that
the details of  their
internal structure and dynamics are irrelevant
for the effective dynamics of the Macros.

More precise expressions of some of these
ideas were encapsulated in the renormalization
group ({\bf RG}) \cite{RG}  and in the
multigrid ({\bf MG}) method \cite{mulg}.

MG was offered for the last 20 years as a computational
philosophy to accelerate the computations
arising in various scientific fields.
The idea was to accelerate algorithms by operating on various
scales such that the computations
related with a certain length scale are performed
directly on the objects relevant at that scale.

In the present review, the multi-scale phenomena
and the relevant Macros hierarchies
are considered for their own interest even (sometimes)
in the absence of a multi-scale algorithm which accelerates
the computations.

The multi-scale concept is proposed as a tool to
reformulate and reorganize the way of approaching the problematics
of various fields.
Thus its usefulness
transcends by far the mere application of a computational
technique and may induce in certain fields a
shift in the concepts, the language, the techniques
and even in their objectives (see chap. 4).

In the last decades, the MicRep
method was used in the RG formulation of the
quantum field gauge theories to explain the
observed behavior of the electromagnetic,
nuclear, radioactive and gravitational phenomena \cite{creutz1}.
One started by representing the continuous space
and time as a discrete
mesh ({\bf lattice}) and representing the field by elementary
simple objects residing on the nodes and links of the mesh.
Each elementary object interacts
by simple microscopic laws
with its immediate neighbors  on the mesh (see chap. 3).

In spite of the fact that the
fundamental interactions act only between
immediate neighbors,
the resulting dynamics of the macroscopic system
is such that the effects of acting on the system
in a particular point are significantly felt
over a large distance $\xi$ around that point.

This phenomenon by which a system defined by microscopic
laws presents macroscopic phenomena is called
"criticality" and $\xi$ is the  "critical length".

The macroscopic dynamics of the
critical systems, at scales larger than $\xi$
is "universal" i.e. largely independent on the details of the
microscopic dynamics used in the microscopic definition
of the system \cite{kadan}.
In particular, universality allows
one to choose a highly simplified
version of the microscopic definition of the system
and still obtain an accurate description of its
macroscopic properties.

Moreover, it was shown that the resulting macroscopic
properties of the critical system can be appropriately
described in terms of the effective dynamics of a
limited number of macroscopic objects (e.g.
"relevant operators" \cite{wils}).

Understanding the critical behavior of a microscopic system
of Micro's
reduces then to the identification of the relevant Macro's
and the description of their long time-scales
evolution.
Conversely, finding an appropriate MicRep
explanation for a macroscopic complex phenomenon is to find
a system of Micro's whose effective macroscopic critical dynamics
leads to Macro's modeling well the macroscopic complex phenomenon.

In practice the iterative algorithms
used to simulate the critical systems in terms of their
simple microscopic laws required exceedingly
long computer runs.
This phenomenon was called
{\bf C}ritical {\bf S}lowing {\bf D}own ({\bf CSD})
(see chap. 2).
A physical way to understand CSD is to realize that
the relevant effective macroscopic processes governing the
Macro's involve typically a macroscopic number of
elementary interactions between each typical pair of
neighboring Micro's.
In fact CSD is a well known, experimentally measurable
phenomenon in the physics of real critical systems
\cite{h}.

Nonetheless its presence in the computer simulations
is a major inconvenient and
a lot of research was invested in the last decade to
devise algorithms which give the
properties of the critical systems without investing the
enormous computer time implied by the presence of CSD \cite{5}.

The construction and study of such  accelerated
{\bf no-CSD} or {\bf reduced-CSD} (RCSD from now on) algorithms,
lead to a finer understanding of the relation
between CSD and the mechanisms which generally
relate the simple microscopic definition of a
system to its complex macroscopic dynamics.

For a system defined by given Micro's,
the RCSD algorithms were based on
identifying and acting upon the relevant Macro's.
As such, these algorithms were at one and the same time
\begin{itemize}
\item the result,
\item the expression
and
\item the source
\end{itemize}
of understanding the critical
dynamics of the systems to which they applied.

In chapter 3 we suggest a quantitative measure
for the "knowledge" or "understanding"
contained in an algorithm.

The emergence of this sort of algorithmic
operational way of acquiring and expressing knowledge
has a very far reaching methodological potential.
 This is even more evident when coupled with the
modern computers simulation, visualization and animation
capabilities \cite{computercenter,viscre}.
 In fact, we performed many  such experiments \cite{computercenter}
(see following chapters) in which
certain systems were successfully studied by
performing computer experiments which involved the
animation of their evolution under the  action of
various algorithms.
The following research routine was common for
most of these experiments:
\begin{itemize}
\item modeling the system as composite of many Micros.
\item computer simulation and visualization of the resulting model.
\item identification and computer experimental study
of the CSD modes within the microscopic system.
\item identification of the Macros.
\item predictions based on the Macros behavior of the
model.
\item comparison with the experimental macroscopic behavior
\item confirming or correcting the microscopic model in view of the
comparison.
\item starting new experiments based on these results.
\end{itemize}

The use of this dialogue with
an artificial  system created in the computer
in order to understand its critical properties
extends to systems
away from equilibrium and to complex systems
which are not
characterized by an energy functional
or by an asymptotic probability distribution.

For such general systems,
some of the notions usually associated to criticality might
become inapplicable.
Yet, one needs criticality in order to insure the
emergence of the universality properties
which make the MicRep method reliable at macroscopic
scales.

We propose to use the
very emergence of CSD in the dynamics
as a criterion for the legitimate use of
universality.
We turn the tables and transform the CSD from
a curse into a blessing in disguise.
We use CSD as the label which isolates the
relevant Macro's.
We use the RCSD algorithms as operational
proofs that the relevant Macro's were
efficiently identified and expressed algorithmically.
We use the Macro's in order to visualize and
understand the emergence of the collective dynamics,
in order to relate the salient complex phenomenology
to the simple underlying microscopic causes
\cite{computercenter}.

We propose to use this understanding
in the task of formulating and studying MicRep
models for basic problems in a wide
range of fields (as detailed in chap. 4).
We are claiming that CSD is the key to characterize,
build, identify and study such models.

\section{CSD Appearance and Measurement}

Let us give a short qualitative description
of the appearance of CSD (and other critical features)
in various systems.
We explain below the fact that
in many cases no fine tuning is necessary
and that CSD is a widely encountered
(in fact typical) phenomenon in nature.

This will become important later in chap. 4 where we advocate the
universal application of the MicRep technique to a very wide
range of complex systems
(For CSD in the graph coloring
problem see Hogg's review in this volume).

One of the typical mechanisms leading to CSD is the following.
At low enough energies, the systems get ordered
and many of their degrees of freedom become effectively
frozen.
The freezing usually results
into a breaking of one of the symmetries of the
system (the spontaneous symmetry breaking phenomenon).

I.e., while the fundamental laws describing the system's
dynamics are symmetric under certain transformations
(e.g. global rotation in the ferromagnetic case,
translation in the crystal case),
the low energy states tend to pick up
particular directions (or positionality) as preferential
(e.g. the direction of magnetization, or the
particular position in which the molecules become localized
upon freezing).

One can show that in this situations,
the long time dynamics of the system is dominated by
the dynamics of large homogenous sub-sets
changing their shape and position collectively
(e.g. spin waves, crystal deformations,
diffusion, Weiss domains, fluid droplets, membrane deformations,
inverse laplacian etc.).

For continuous symmetries and local interactions,
one can explicitly
identify in these situations long(-macroscopic)-range
correlations in space and time (Goldstone modes-
e.g. long-wavelength spin waves in isotropic
magnets) \cite{gold}).

As opposed to usual critical behavior which
takes place only for very special (fine-tuned)
values of the parameters of the model
(e.g. critical temperature $T_c$), the
long range excitations related to the
Goldstone mechanism take place naturally for
ranges of non-zero measure in the parameters
space defining the systems (e.g. $T < T_c$).

More generally, if a system presents
states which differ slightly in their energy
from the ground state, one can have a dynamics
which interpolates between the various (quasi-)vacua and
which takes macroscopic times (e.g spin glasses \cite{spin}).

In general, this is not necessarily related to the breaking of
a  global symmetry.
All it is necessary is that there exist  collective changes
over a  large  subset of the system which
do not  result into  a significant change of the energy per
degree of freedom \cite{parisi}.
Even though large numbers of elementary degrees
of freedom are involved,
it is therefore not possible in general  to
expose long range correlation lengths
within the system\footnote{
Another such instance where criticality is not related
to long range space-correlations is
the completely connected Ising model
where all the sites are at distance 1
one from the other and still criticality and CSD
is  present for "local" (or rather "single-spin") dynamics.
The proper terminology would be therefore
"large sub-sets correlations" and the RCSD dynamics
may be called "large sub-sets dynamics".}.
It is  therefore important that
even in the absence of a divergent $\xi$, the CSD phenomenon
is still there to  signal that
the system is still critical and universality may
still apply.

Moreover, criticality and universality labeled by CSD
apply to systems far from the equilibrium.
For such systems, CSD may just be the result
of "branching": i.e. the existence of alternative courses of the
far-from equilibrium dynamics
which are equally likely to take place.
When followed in time (or other parameter)
the different courses of evolution  might be separated by
very small energy barriers at the beginning and by
very large barriers in the end\footnote{
In our examples in the next sections,
such dynamical branching
phenomena happen when one freezes a system with instantons
and gets trapped in various topological sectors,
or when the traffic flow chooses between organizing into an optimal
flow or into a completely jammed configuration.}.
This makes the branching region analogous
to a phase transitions in equilibrium systems
and the situation after branching analogous to the
"magnetized phase" where the system has degenerate
"free energy" minima and the dynamics breaks the
symmetry which relates these minima by effectively
choosing one of them.

Note that as one lowers the available thermal energy
and the tunneling between vacuums becomes
rarer and rarer, the corresponding tunneling time scale
$\tau$ does not shrink. In fact $\tau$ diverges
to higher and higher values until
the time scales involved become longer than
the typical runs can afford.
For a finite system and non-zero energy, there is always a time
above which the tunneling still can take place.

As a consequence, the Critical Slowing Down regime
is not limited to a certain value of the parameters
(e.g. for temperature $T_c$), but to all values of parameters
for which the  depth of the (local) minima
in the "energy landscape"
are larger than the typical thermal energy available within
the system at typical times and locations \cite{parisi}.

In this sense, the CSD once installed
never disappears as the  available
energy gets smaller and smaller:
the usual critical point is not any more
the singular point at which time correlations
diverge, but in fact the point at which
the CSD related to (quasi-)ground states
degeneracy DISAPPEARS(!)
as the temperature rises above the
energy landscape.

Another effect which favorites the presence in nature
of critical and CSD systems
and
gives them a non-zero probability measure to appear
is the fact that once a system is getting
into a critical state, the very presence of CSD
is preventing it to leave this state within a microscopic time.

For instance if one lets water to cool from $200^0 C$
to $-100^0C$, the water will spend a non-zero fraction of time at
exactly $100^0C$ and respectively $0^0C$
(in this particular case the transition is first order,
so it is not clear
how much   of the macroscopic properties are
independent on the microscopic details).

Also, if one is holding different boundaries of a water
container at respectively $90^0C$ and $110^0C$, a macroscopic
fraction of the volume of the container will present
the bubbling and large fluctuations characteristic
to the critical temperature $100^0C$.

We will insist below on this
relation between the emergence of objects
with macroscopic space dimensions and the emergence of
processes extending over macroscopic times.

This property holds for non-equilibrium systems too
(actually $especially$ for non-equilibrium systems).
In both equilibrium and non-equilibrium systems,
the system spends more time and/or more space
in the critical region, in a $natural$ way, without the
necessity of fine tuning of the boundary
or of the initial conditions.

It is worth mentioning that from the simulation point
of view all the past runs  aimed at
bringing statistical mechanics and lattice field theory models
towards the thermal equilibrium state,
become particular instances in which
the CSD -labeled non-equilibrium universal dynamics
was studied and characterized
(2D XY, roughening, Weiss domains,
Dirac equation in gauge background etc).

Another general CSD generating mechanism
further blurring the difference
between   static and dynamic simulations is
the  "decoupling" mechanism of certain
(metastable) global degrees of freedom \cite{fuc}:
if the freezing of the system takes place
much faster than the typical times necessary for
certain global structures to develop or disappear,
the system gets "quenched" in configurations
which are not typical for the thermal equilibrium
at the frozen temperature.
Indeed, in as far as the short range interactions
are concerned, the system is at the freezing temperature
but the large distance features are still typical
for the higher temperature.
By further lowering the temperature, the
relaxation of the long-distance features
towards the low energy configurations becomes
$slower$ rather than being accelerated.
In particular the relaxation dynamics of the
global features is not directly
related to the various time scales of the
external factors which pump or
extract energy from the system.
The various space scales of the global quenched
objects are the ones which govern the
corresponding multiscale hierarchy
of relaxation times.
The CSD becomes consequently the repository
of the information related with the
relevant Macros of the system.

In order to clarify quantitatively the relation
between space scales and time scales, we devised
a general tool: the Multi-Scale Dynamical Experiment
(MSDE) \cite{nathan}.
While for MSDE calibration, we used initially
an equilibrium experiment, its main destination
is in the questions of nonequilibrium dynamics mentioned above.

Following \cite{bb},
we defined the elements of the space-time-correlation matrix to be:
\begin{equation}
M_{\bf{l}, \bf{j}}(t)=
<S_{\bf{l}}(0)S_{\bf{j}}(t)>-<S_{\bf{l}}><S_{\bf{j}}>.
\end{equation}
where $\bf{l}$ and $\bf{j}$ are points on the lattice
(e.g in the square lattice
${\bf l}=(n_x,n_y), ~ 1\leq n_x \leq L_x, 1\leq n_y \leq L_y$ ),
and $S_{\bf{l}}$
is any local observable (e.g. the spin at the point $\bf{l}$).
$t$ is the (running) time between the measurement of  $S_{\bf{l}}$
and the measurement of $S_{\bf{j}}$.

In the case of a translation invariant dynamics
(e.g. as defined by the eq. (9)) one has:
\begin{equation}
M_{\bf{l}+\bf{d},\bf{j}+\bf{d}}(t)=M_{\bf{l},\bf{j}}(t)
\end{equation}
where $\bf{d}$ is any lattice vector.
Thus, one can identify the eigenvectors of the
space-time-correlation matrix as Fourier modes. Each
eigenvalue $\lambda_{\bf{k}}(t) $ corresponding to the
eigenvector $ e^{i \bf{j} \cdot \bf{k}}$ satisfies:
\begin{equation}
\sum_{\bf{j}} M_{\bf{l},\bf{j}}(t) e^{i \bf{j} \cdot \bf{k}}=
\lambda_{\bf{k}}(t) e^{i \bf{l} \cdot \bf{k}}.
\end{equation}
 For each wave-vector $\bf{k}$ one can assign a decorrelation time
 $\tau_{\bf{k}}$ via:
\begin{equation}
\lambda_{\bf{k}}(t)\sim e^{-t/\tau_{\bf{k}}}, ~~~~~~
 t\rightarrow\infty
\end{equation}
Note that for instance in the spin case,
$\tau_{\bf 0}$ ($\tau$ for ${\bf k}=(0,0))$
is identical to the usually defined
exponential correlation time for the magnetization.

Measuring the decay of the eigenvalue, $\lambda_{\bf{k}}(t) $,
we estimate $z$ assuming the dynamical RG \cite{ma} scaling relation:
\begin{equation}
 \tau_{k}\propto k^{-z}.
\end{equation}
where $k$ denotes the norm of $\bf{k}$.
The reason for this assumption is that the corresponding length scale
of a Fourier mode is:
$\Lambda\propto k^{-1} $. Therefore assuming dynamical universality
one expects that $ \tau_k\propto \Lambda^{z}$.

\em
	 The Simulation.\rm
{\Large\bf$-$}The model in which we calibrated our method is
the 2D Ising on a square lattice \cite{IS}.
We measured the correlation eigenvalues for both
a global algorithm ( Wolff  \cite{W})
and a local one (Parallel Checkerboard Metropolis (PCM) dynamics).
We use 2 lattice sizes $L=64$ and $L= 128$
to show that the result is independent in L.

We worked exactly at the critical temperature
$T=T_c={2/{log(\sqrt2+1)}}$ (we normalize $J$ and $k$ to $1$).
In the PCM case we employed the Wolff dynamics
to decorrelate the system between the "local" PCM episodes
when we performed the measurements.
In both cases we had periodic boundary conditions.
In both cases we discard the first 10,000 Wolff sweeps and then
took $3\times10^5-2\times10^6$ different measurements of the
space-time-correlation matrix, with ten Wolff
sweeps between each measurement.
Due to the translation invariance it is enough to
compute only one line of the space-time-correlation
matrix in each dimension, and the reflection symmetry reduces
further the required measurements to almost
half of a line in each dimension, altogether only $ (L/4+1)^2$ elements.
We repeated for 10 independent systems and estimated the
error bars by the standard deviation.
The fitting to an exponential,
in order to estimate $\tau_{\bf k}$, and to power law
to deduce $z$, were done using nonlinear least square fitting \cite{9}
which gives the error bars as well.

In figures 1-3 one can see a log-log plot of
$\tau_{\bf{k}}(k)$, and the evaluated
$z$ for the Metropolis and Wolff dynamics, respectively
(we ignored the smallest $k$ values where the linear fit fails
because the saturation of the time correlations due
to finite size effects.
In particular, in MSDE,
the usual measurement of the total magnetization is
a datum which we discard as a mere finite size artifact.

The final results of our fitting are
$z_{\rm{met}}=2.1\pm0.05$, $z_{\rm{wolff}}=0.1\pm0.2$,
for L=64,
and
$z_{\rm{met}}=2.1\pm0.1$, $z_{\rm{wolff}}=
0\pm0.15$,
for L=128.

MSDE display explicitly the \em simultaneous \rm existence
of \em all \rm the time scales below the maximal scale in the system.
For increasingly larger systems, this scale diverges
and the critical system presents explicitly
all the time scales as active scales in its dynamics.
This produces in the infinite volume limit a power law.
The MSDE  method investigates the internal relations
between the spatial and the time scales in statistical models
and translates into practical numerical experiments
the ideas of the dynamical RG.

The results in the 2D Ising model confirmed
that the Dynamical RG assumption for the scaling relation
$ \tau_{\Lambda} \propto \Lambda^{z}$
holds both for the Metropolis and the Wolff dynamics,
suggesting that it does not depend on the particular kind of
dynamics.

One can go further and use our method to analyze the
internal structure of the Fourier subspaces as well
as seeing the correspondence between
the equilibrium and non-equilibrium measurements.

The experiments relating the spatial scales and features with the
time scales of CSD of the type reported above
can help us identify the relevant  macros in other models.
In particular the ferromagnetic Ising model on certain lattices
has at very low temperatures, in spite of the unique ground state
some dynamical spin-glass-like properties.

For instance, for Ising spins defined on the triangles
of a Random Triangulated Surface (RTS) \cite{kinar},
the presence of baby universes of arbitrary large
size $L_{baby}$  leads to islands of uniform spin
whose life-time diverge like $\tau_{baby} \sim e^{L_{baby} / T}$
(i.e. they are rigorously stable at $T=0$)
\footnote
{Even on a finite regular hexagonal lattice the  life times of
the convex  islands of linear size $L$
are of the order $\tau_{convex} \sim L^3 e^{1/T}$
and therefore they become rigorously stable at $T=0$.}.

The natural branched polymer structure of both closed \cite{agi}
and open \cite{pazi,esti} RTS
gives rise to an ultrametric structure
of these states landscape which is very similar with the
one typical for spin glasses \cite{parisi}.

It is encouraging that even in such conditions algorithms
acting on large spin islands
are effective in eliminating CSD \cite{kinar}.
The Ising model on an infinite coordination number lattice
is also tractable by the cluster algorithms \cite{idoradi}.
For low average coordination number, the
global cluster algorithms
may apply even in the frustrated case \cite{idoetala}.

In both equilibrium and non-equilibrium systems
we have used successfully interactive computer animation
in order to identify the spatial structures and the Macros responsible
for the CSD. In particular we ran in parallel local
(CSD) and global (RCSD)
algorithms and documented the evolution of the systems at various
spatial scales (see last section).

\section{ Elimination of CSD as Understanding of Macroscopic Dynamics}

The perception of Critical Slowing Down as an unwanted feature
of the simulation, lead into oblivion in the past studies
the fundamental importance of CSD as a tool in identifying the
relevant degrees of freedom
of the particular critical system under study.

However, in the context of RCSD algorithms
the fact that the acceleration of
the dynamics of  a certain mode
eliminates/reduces  CSD is a clear sign that the critical degrees of
freedom were correctly identified and their
dynamics correctly understood.

RCSD algorithms express and validate in an objective way
(by reducing the dynamical critical exponent $z$)
our previous intuitive understanding of the collective
macroscopic dynamics of large ensembles of microscopic elementary
objects.
In certain systems, which resist the conceptualization
of their understanding in closed analytic formulae,
this kind of "algorithmic operational understanding"
might be the only alternative.

With this in mind,
one may attempt to use CSD and its
elimination (reduction) as a quantitative
measure of the "understanding" of the model's critical properties.
For instance an algorithm leading to $z=2$
would contain a very low level of understanding
while the "ultimate" level of understanding
 is when one needs not simulate the model at all
in order to get the static properties of every "lattice size"
(analytic understanding).

At the other extreme
are the simulations of spin glasses where
the the correlation times increase exponentially with the size of the
system.

In simplicial
quantum gravity the increase is faster than any function
\cite{nabu}.
This would be the "ultimate" CSD: {\it in}computability;
that is,
$a \ priori$ impossibility to compute
within a systematic approximation procedure
the numbers "predicted" by the theory \cite{nabu}.
The rise of incomputability in the context of the MicRep
approach allows a new perspective on the issues of
predictability  and reductionism:
the possibility arises that the future
state the universe is physically totally $determined$
by its present state,
yet the future state cannot be $predicted$ because its
physical parameters as determined by the theory are
mathematically uncomputable numbers.
(Unfortunately
this fascinating point falls outside the scope of
present article.)

Until now,
one could think that there are only two possibilities:
to know or not to know. However, as explained above,
during the last years research it became evident that intermediate levels
of understanding exist. They are all categorized by
the ability to eliminate or
reduce
the CSD. We have learned that in many models embedding a knowledge that
we have about the model can result in a better (faster) dynamics.

Often, the way to a "more efficient algorithm"
   passed
through the understanding of the "relevant degrees of freedom" and their
dynamics. In order to maintain all the information stored by the elementary
degrees of freedom, it is always necessary to define precisely the relevant
Macros in terms of the Micros.
Once the Macros are defined in term of the Micros it is possible to find their
dynamics in term of the translated Micro dynamics.

There are few "tests"
to establish that for a given critical system the set of
Macro's was correctly identified:
\begin{itemize}
\item
One would need to make sure that there is a "large" number of Macro's.
This requirement makes sure that a large fraction of the relevant degrees
of freedom is indeed represented by the Macro's that were discovered.
\item
Then one has to check that they are relevant in the sense that they are not
just symmetries of the theory. In other words, a change in an Macro should
have an influence on the important measurables of the system.
\item
One of the more stringent tests is to
verify that the resulting Macro dynamics is "free". That is, in a typical
configuration of Micro's the resulting dynamics of the Macro seems to be
free. This is a signal that the correct Macro have been identified.
\end{itemize}
An analogy for the relation between Macro's and Micro's can be
found in language. The letters would be the Micro's and the words will be
the Macro's.
Of course, manipulating words amounts to manipulating letters.
However when one "works"  in the words-level one need not bother
with the letter-level, even though these two levels co-exist.

In order to illustrate the way in which one
can eliminate CSD by identifying and acting upon
the appropriate Macros, let us consider the case
of a free field on a Random Triangulated Surface (RTS).

In particular this example illustrates simply
the relation between the rules constructing the
Macros and the critical properties of the physical model.

In the example below the macros overlap
rather than being separated by sharp boundaries.
In fact, the same Micro may belong (to various degrees)
to more than one Macro. This "fuzziness"
rends the boundaries defining a Macro
a "subjective" choice , a matter of degree/opinion,
which, while not affecting the validity of
the numerical algorithm,
sets the scene for further conceptual elasticity.
 It suggests continuous interpolation and extrapolation
procedures closer in their form and essence
to the working of the natural human intelligence.

In fact, through substituting the binary logic
of the Micros with the continuous one of the Macros,
one may avoid the no-go theorems, the paradoxes
and the puzzles related to
({\it un})computability, ({\it i})reversibility and
({\it creative}) reasoning.

The precise yet "smeared" formulation of the Macros
within the multiscale MicRep approach
bypasses these classical conceptual puzzles arising
in the reductionist method. In particular, while the Macros
acquire a certain degree of reality,  individuality
and causal behavior at the macroscopic level,
their conceptual boundaries are fuzzy enough to avoid
the paradoxes which would arise should one try to apply
Macro categories in the microscopic domain
(their boundaries "dissolve" gracefully as one tries to
resolve them with a too sharp "microscope").
In the MicRep Multiscale framework, there is no contradiction between
considering the ocean as a set of molecules or a mass of waves.
These are just complementary pictures relevant at different scales.

In the theory of  random triangulations \cite{triang}
one has usually to calculate
expectation values, $< O >$, of observables, $O$, according to the formula:
\begin{equation}\label{expect}
		  < O > = \sum_{\{X\}} O(\{X\})P(\{X\})
\end{equation}
where the sum is over all field configurations, $\{X\}$, and
\begin{eqnarray}
  P(\{X\}) = \frac{1}{Z} \exp (-\beta E(\{X\})  \\
\nonumber Z = \sum_{\{X\}} \exp (-\beta E(\{X\})
\end{eqnarray}

The expression for the energy,
in presence of an external field, $H_{i}$ is:
\begin{equation}\label{energy}
		 E(\{X\}) = \sum_{<i,j>}J_{ij}x_{i}x_{j} + \sum_{i} H_{i}x_{i} + E_{0},
\end{equation}
where the first summation is over all the links,
and $x_i$ is the field on the site $i$.
In the  free field case $J_{ij} = -1$, and
$J_{ii} = 3+m^2$ where $m$ is the bare mass.

Note that on a lattice with arbitrary topology
(which will become our case after the first coarsening
step), for an arbitrary
number of neighbors $N_i$ of the point $i$,
the relation between the mass $m$ and the coefficients
is $J_{ii} = N_i + m^2$ and $ J_{ij} = -1 $.
Therefore the signal that a particular link
is susceptible of being part of the mechanism
of creating long range correlations ($m \approx 0 $) and CSD
is $J_{ij} \approx -{ {J_{ii}}/ {N_i} }$
while links for which $ -J_{ij} << { {J_{ii} }/ {N_i} }$
are allowing each of their end nodes to equilibrate quite
independently one from the other ($m >> 0$).
Therefore they are not significant for long range effects
and can be effectively treated by local algorithms.
This will become important later in deciding the blocking procedure
for the global algorithm.

The common local
 method to estimate (\ref{expect}) is the heat-bath (HB)
\cite{binder} algorithm.
It
creates iteratively configurations according to the following procedure.
Given a configuration $\{X\}$ one constructs a new configuration $\{X'\}$
by updating the field $x_{i_{0}}$ at an arbitrary point $i_{0}$.
More precisely, $x_{i_{0}}$ will be changed to $ x'_{i_{0}}$
with a probability,
\begin{eqnarray}
	 W(\{X\} \rightarrow \{X'\}) =            \\
\nonumber	 =
N \exp ( - \sum_{j} J_{i_{0}j} x'_{i_{0}}x_{j} + H_{i_{0}}x'_{i_{0}}),
\end{eqnarray}

 where $N$ is a normalization factor.
 That probability
 fulfills the detailed balance condition \cite{binder},
\begin{eqnarray} \label{detbal}
		W(\{X\} \rightarrow \{X'\})P(\{X\}) =   \\
\nonumber		=W(\{X'\} \rightarrow \{X\})P(\{X'\}).
\end{eqnarray}

The condition (\ref{detbal}) is sufficient for taking care of the
factor $P(\{X\})$ in estimating (\ref{expect}).
I.e., after creating n decorrelated
configurations
according to (\ref{detbal})
, one obtains
(\ref{expect}) by simply computing the value of the observable
$O$ for each of those configurations
and taking their average:
\begin{equation}\label{meanvalue}
		< O > \approx \frac{1}{n} \sum_{\{X\}} O(\{X\}).
\end{equation}

Unfortunately, HB generates strongly time-correlated
configurations as the
number of the grid points increases.
This critical slowing down (CSD), related
with fields correlations,
is well known \cite{h} and analog to the CSD of laplacian-like
Partial Differential Equations (PDE) solvers (e.g. Gauss-Seidel ).
Therefore, in constructing our algorithm,
we borrow intuition from the non-dynamical PDE solvers
terminology.

In particular, the process of eliminating the errors
during the iterations of an Algebraic Multigrid (AMG)
PDE solver is analogous  to the  elimination
of correlations in the present  Dynamical
Algebraic Multigrid (DAMG) procedure
(and to reaching the thermal equilibrium
at the beginning of the runs).

The fact that this conceptual
analogy works is of course a  non trivial
statement on the critical dynamics of the system.
Consequently, the success of our procedure
expresses a nontrivial understanding (which turns
out to be of degree $z=0.2 \pm 0.2$).

As in the non-dynamical AMG solvers case,
we define first an $\alpha-$strong connection. A point $i_{0}$
is $\alpha-$strongly connected to $j_{0}$, if
\begin{equation}
 J_{i_{0}j_{0}} > \alpha \ . \ max \{J_{i_{0}k} \}
\ \ \ \ \ \ 0 \leq \alpha \leq 1,
\end{equation}

In the PDE solver case, the CSD originates in the fact that,
after some relaxation sweeps, the error  becomes smooth
over strongly connected sets of points.
While the smooth components of the error are not reduced efficiently
by the usual relaxation, they are easily transported to the coarse grid
where they can be efficiently reduced.
Thus, coarsening, in the AMG scheme is done by
blocking strongly connected points.

The strong connection link criterion is blocking together
variables which interact via links corresponding
to low mass contribution to the local energy
 (according to the comments following equation 8).
These are the links which are most difficult to equilibrate
by local updatings and which are responsible for long range
correlations and CSD.
One sees again how the building of a RCSD algorithm
is in fact the expression of a more intimate
understanding of the relation
between the microscopically defined elementary interactions
and their macroscopic consequences.

 The connections $J_{ij}$ appear also as weights in expressing
the values of functions (and operators) on the coarse grid points
as weighted averages
of the values on the fine grid points.
It is interesting therefore that even though the model is definitely not
a gauge model, its multiscale structure endows the $J_{ij}$ coefficients
with parallel transporters properties.

We have constructed the coarse grid according to the
prescription developed for PDE  in \cite{amg}.

However, in dynamical relaxations, utmost care has to be taken to fulfill
the detailed balance condition, (\ref{detbal}).
This puts additional constraints on the way the transport
of the variables (and operators) between
the grids is performed.
To see it, consider the change in the coarse grid energy caused by
the updating of a coarse grid variable.
When translated in terms of the fine grid variables, this
coarse grid  updating leads to a change in the fine grid energy.
We will show now that
the detailed balance condition is fulfilled
if this change in the fine grid
energy is equal to the change in the coarse grid energy.

We will show it for a two-levels cycle as
the similar property for multilevel cycles follows
from iterating the two level argument.

The starting point is a fine grid configuration, $\{X\}$,
obtained after some fine grid HB relaxation sweeps.
The variables are
$x_{i}$
and their current values,
are $ x^{0}_{i} $.
The aim is to update these values using a two levels cycle.

 In the case of (non-dynamical)
 MG  solvers what is transported on the
coarser level is not the fine grid  equation but the
so called "error equation" \cite{mulg}.
By analogy, in the dynamical case (DAMG) we consider
the following change of variables:
\begin{equation} \label{qi}
x_{i} \rightarrow x^{0}_{i} + q_i.
\end{equation}
(the new fine grid variables,
$q_i$, are analog to the fine grid error components in the non-dynamical
algorithms).
Every change in the new variables $q_i$
expresses a corresponding updating of the
variables $x_i$.
 The variables $q_i$ are defining a configuration, $\{Q\}$.
By substituting (\ref{qi})  to (\ref{energy}) the
energy is expressed in terms of $q$'s as:
\begin{eqnarray} \label{qenergy}
         E(\{Q\})  =
\sum_{i,j} J_{ij} q_{i}q_{j} + \sum_{i} H_i'q_i + E_{0}',\\
\nonumber H_{i}' = H_i + 2\sum_j J_{ij}  x_{i}^{0},\\
\nonumber E_{0}' =E_{0} + \sum_{i}H_{i}x_{i}^{0} + \sum_{i,j}  x_i^{0} x_j^{0}
\end{eqnarray}
Note that even when starting with $H_{i}=0$
in (\ref{energy}), non vanishing external fields $H_i^\prime$
are arising
in (\ref{qenergy}).

The next step in the usual, (non-dynamical) AMG solver is to
transport the configuration $\{Q\}$ to the coarse grid in
order to accelerate the updating.
In the dynamical case the coarsening
corresponds to considering during a coarse sweep
only a selected type
of updatings, parameterized by a smaller number of
parameters $q^c$ than the number of degrees of freedom $q$.

More precisely in DAMG
we are considering only updatings which affect the
following  nonlocal combinations of fine grid variables:
\begin{equation}\label{continuum}
         q_{i} = I_{ij} q^{c}_{j},
\end{equation}
where in eq.~(\ref{continuum})  $I_{ij}$ is the (rectangular)
 interpolation matrix,
and the  $ q^{c}_{j} $ can be called "coarse grid variables".

Since updating the $q^c$'s does not lead in general
from every fine grid configuration to any other arbitrary
fine grid configuration,
the updatings parameterized by changes in the
$q^c$'s
forsake  ergodicity. In exchange, these nonlocal steps
are more efficient in traveling throughout the configurations
space.
The ergodicity is restored
by the performance of usual sweeps
 on the fine grid.

The correct Boltzmann distribution  is insured
by taking care that the configurations
 as expressed by  various values of $q^c$
are distributed according to their energy when expressed
in terms of the fine grid variables $q$.

This means that the distribution of the
variables $q^c$ is governed by the energy (\ref{qenergy})
when expressed in terms of (\ref{continuum}):
\begin{eqnarray}
	E(\{Q_c\}) = \sum_{i,j} J^{c}_{ij} q^{c}_{i}q^{c}_{j}
+ \sum_{i}H^{c}_{i}q^{c}_{i} +E_{0}'  \\
\nonumber	 J^{c} = I^{t}JI,\\
\nonumber    H^{c} = I^{t}H'.
\end{eqnarray}

One can see now that
when restricting the updates of the system to
the coarse moves parameterized by the coarse variables
$q^c$, one is lead to simulate a system of the
same type as the initial one (\ref{qenergy}) with new coefficients
$J^c_{ij}$.
This allows one to proceed iteratively and apply the
coarsening procedure to the system $q^\prime$.

This completes the definition of the DAMG cycle.
The coarsest system, which contains a very small number of variables
and quite short range correlations can be simulated
readily by HB or Metropolis methods without any worry of CSD.

{\em Correlations - The CSD in the dynamical process. }

The objective of the Dynamical Algebraic Multigrid (DAMG)
 algorithm is to eliminate/reduce the CSD.

In order to characterize CSD, one defines first
the autocorrelation function, $C(n)$.
Let $\{X\}_{i}$, $i=1,..,N$, be a chain of configurations
obtained by a Markov  process such as the one produced by the HB procedure.
Let $O(\{ X_i \} )$ be the results of measuring
a certain observable $O$ on each of the configurations.
 The auto-correlation function is defined by:
\begin{equation}\label{cor1}
         C(n) = <O(\{X\}_{i}) O(\{X\}_{i+n})> - <O>^2.
\end{equation}
At large n, $ C(n)$ behaves like,
\begin{equation}\label{cor2}
         C(n) \sim C(0) \exp{(-n/\tau)},
\end{equation}
where $\tau$ is the so called auto-correlation 'time'.

$\tau$ measures the number of Markov steps one has
to perform after obtaining a configuration in order to obtain
a new configuration uncorrelated with the first one.

In critical systems, $\tau$ diverges with the dimension of the system.
In the case of a regular $d$-dimensional grids of volume $L^d$, one has:
\begin{equation}
\tau \sim L^z.
\end{equation}
which defines  the dynamical critical exponent
$z$. Typically $z \geq 2.$ for local algorithms.
In our case, since the lattice is not regular,
we checked the dependence of $\tau$ on the number of
triangles $N$:

\begin{equation}\label{zeta}
\tau \sim N^{\zeta},
\end{equation}
which assuming an effective
(Hausdorff) dimension $d_{H}$ of the
lattice would be related to $z$ by:
\begin{equation}\label{eta1}
z = \zeta  d_H .
\end{equation}

For a typical 2-dimensional critical system which displays
CSD, $\zeta$ would be of order $1.$

In the next section we show the comparison between the
$\zeta$ value in DAMG runs vs. usual one level HB.

{\em Results}

The critical properties of the system
are best represented by the surface "width":
\begin{equation}\label{fluc}
O=<(x-<x>)^2>
\end{equation}
which in the string formalism is also called
the mean square extent of the surface.

This quantity diverges with the size of the system
and consequently displays CSD during local runs.
Thus, we have measured $O$-related quantities, both
using DAMG and using the  usual one level HB for comparison.

In Figure 4 we demonstrate CSD
by showing that the local HB run has problems in reaching
the correct value within a reasonable time
and has extremely long range correlations.
As opposed to it, the DAMG has correlation times of the
order 1 and covers very effectively the configuration space.

In order to make a fair comparison one has to take into account
that a DAMG sweep involves more operations than a HB one.
Consequently, we
compared the computations only after normalizing the
units such that one unit represents the same CPU time in both cases.
In practice, this reduces to counting the number of HB 'hits'
performed during the run (a 'hit' is
a local updating of a field upon a site) . In  figure 4,
the ordinate $n$ is in units of the number of hits performed upon the
fine grid, that is
\begin{equation} \label{norm} n= (number\ of\ hits \ performed) /{N} .
\end{equation}

The lattice size of the run presented in Fig. 4
was $N= 8596$ triangles and
one level measurements were made after each 'sweep',
i.e. $N$ 'hits'. DAMG measurements were made after each 'cycle'.
A 'cycle' contains
a sweep upon the fine grid and all the coarser grids sweeps needed
to update it.

Figure 5 is a comparison of the autocorrelation times, $\tau$,
for DAMG vs. one level HB.
The grid size was $N=10000$, and using the formula (\ref{cor1})
we have computed the logarithm of correlation function, $C(n)/C(0)$,
for $O$.
Here we have also normalized the running time
in the way mentioned above,
in order to have a common basis of comparison between the
two methods.

We measured the slope of $\log C$ and got (see (\ref{cor2}):
$\tau$ = 1661 (143) for the one level HB,
and
$\tau$ = 2.13 (0.24),
in the DAMG scheme
where the numbers in parenthesis are the errors.

Figure 6 presents the autocorrelation time, $\tau$,
as
a function of the number of grid points, $N$.
{}From (\ref{zeta}) one can see that plotting it on a $\log - \log$
scale, and measuring its slope would give the critical exponent $\zeta$.
The results for usual HB is
$ \zeta = 1.5 (0.2)$ while for DAMG $\zeta= 0.2(0.2)$.

Let us emphasize that beyond the
computational gain of eliminating CSD completely,
 we obtained here nontrivial physical information
on the character and the origins
of the criticality in this model.
In particular we learned about the relation
between the microscopic "strong links"
and the macroscopic long range and long time correlations.
In this simple system, this can be traced to the
relation between the microscopic bare mass $m$ appearing
in the definition of  link energy and the
macroscopic correlation length corresponding
to the physical mass:
$m_{phys} \sim {1 \over {\xi}}$ \cite{DDK}.

The Dynamical Algebraic Multi-Grid
for free fields on RTS is just an example of the general
computational philosophy we applied in a full range of models.
The very high precision physical measurements
obtained with this new generation
of efficient algorithms were reported elsewhere
 \cite{elsewhere}.

We summarize below some of those applications
by providing for each of them
a relevant reference, the Micros, the elementary INTERactions
and the relevant Macros.
According to the description above,
for real fields on RTS this summary has the form:

\begin{itemize}
\item Elimination of CSD for free fields simulations on arbitrary lattices
 by acting on the relevant collective variables \cite{wolo}.
         \begin{itemize}
          \item Micro  - value of field at a point
          \item INTER  - continuity on strong links, noise on weak links
          \item Macro  - large neighborhoods with strongly correlated fields.
         \end{itemize}
\end{itemize}

For other systems, we have:
\begin {itemize}
\item Elimination of CSD by performing the relevant critical operations.
      The example of Valleys-to-Mountains-Reflection (VMR) algorithm
      in interface roughening dynamics \cite{evertz2}.
         \begin{itemize}
          \item Micro  -the height of surface at certain point
          \item INTER  - height difference
          \item Macro  - valleys and mountains
         \end{itemize}
\item Elimination of CSD by acting on the relevant subsystems.
      The example of Ising spins and of discrete
      gaussian model on regular and RT Surfaces \cite{kinar}.
         \begin{itemize}
          \item Micro  - sign at a point
          \item INTER  - preference for the same sign neighbors
          \item Macro  - blocks of same sign
         \end{itemize}
\item  Eliminating or reduction
of CSD by acting on macroscopic Gauge invariant degrees
   of freedom :
\begin{itemize}
       \item  Flux Tubes Dynamics \cite{z2}.
         \begin{itemize}
          \item Micro  - gauge fields associated to links
          \item INTER  - plaquette frustration
          \item Macro  - flux tubes
         \end{itemize}
       \item  Instanton Dynamics \cite{longPTMG}.
         \begin{itemize}
          \item Micro  - gauge fields on links
          \item INTER  - plaquette frustration
          \item Macro  - topologically stable instantons
         \end{itemize}
       \item  Polyakov Loop Dynamics \cite{poly}.
         \begin{itemize}
          \item Micro  - gauge fields on links
          \item INTER  - plaquette frustration
          \item Macro  - polyakov loop aligned blocks
         \end{itemize}
       \item  Fermion Dynamics in Gauge background
         and Parallel Transported Multigrid \cite{visu}.
         \begin{itemize}
          \item Micro  - fermion fields
          \item INTER  - local Dirac equation
          \item Macro  - extended (topological) solutions to Dirac equation
         \end{itemize}
\end{itemize}
\item  Elimination of CSD in the geometry dynamics of 2D SQG \cite{agi}.
         \begin{itemize}
          \item Micro  -points, triangles
          \item INTER  - connectivity
          \item Macro  - global geometry, topology.
         \end{itemize}

\end{itemize}

Another phenomenon which appeared in the Valley-to-Mountain
 algorithm applications
in the application of Instanton global moves and
in the finite temperature polyakov loop application
is the "hetergodicity".
This term, denotes a situation in which the juxtaposition of steps
belonging to two slow algorithms leads to an
algorithm which does not present critical slowing down
\cite{evertz1,longPTMG,poly}.
This turned out to give precious information
on the structure of the critical modes subspace and its
expression at the Markov matrix level \cite{longPTMG}.

\section{MicRep use in Multiscale Phenomena}

As explained in the previous sections,
we propose to use the experience gathered in
isolating the relevant macroscopic dynamics in order
to study various complex models not necessarily at equilibrium.

A context in which the MicRep is unavoidable, is when
the phenomena under study do involve a multitude of scales.

Moreover we have argued in chap. 2 that one may expect
a reasonable degree of universality even in systems
which did not undergo a fine tuning in their parameters and
even in systems which do not admit a rigorous
equilibrium or near-equilibrium Statistical Mechanics
definition.

We sketch here a few examples and then concentrate in full detail
on one of the applications.

{\bf Coarse grained Molecular Dynamics}

Using a very simple minded "Coarse Grained Molecular Dynamics"
{\bf CGMD} method of simulation \cite{zeiri},
we were able to describe and predict
the nonequilibrium distributions of reactants
in certain Chemical Reactions.
The microscopic "elementary" objects did not necessarily represent
the molecules dynamics faithfully:
the "elementary" Micros moved on straight trajectories
except for certain cross sections for elastic
and inelastic collisions as well as a cross section for
the chemical reaction to take place if the colliding
Micros were in the appropriate excited internal state.
In spite of its coarseness,
this approach allowed the realistic treatment of  reaction rates
and chemical distributions arbitrarily far away from equilibrium.

In analogy to the scheme used in  previous chapter, we
may describe this application schematically
as:
\begin{itemize}
   \item {\bf coarse grained molecular dynamics}
     in far from equilibrium systems \cite{zeiri}.
      \begin{itemize}
      \item Micro  - "molecules"
      \item INTER  - collisions, excitation, activation, bonding
                     and dissociation.
      \item Macro  - reaction rate, density and energy distributions.
      \end{itemize}
\end{itemize}

The use of this type of CGMD might be absolutely unavoidable
in describing systems in which various reaction chains
are mutually exclusive and a $de \ facto$ segregation
takes place in the system between regions in which
various mutually-exclusive reactions take place.
We have in mind the situation in which
the same set of enzymes are capable in principle to
sustain 2 or more different reaction chains
but due to certain inhibitory cross-interactions taking place
at certain intermediate steps between the chains,
the chains cannot take place simultaneously in the
same spatial region. There is preliminary
evidence also for the opposite effect:
reaction chains mutually sustaining one
another into active states which would not be
viable in the absence of their coexistence
and cross interaction \cite{bagrat}.

The implicit assumption of continuity imposed by
a continuum hydrodynamic - like or Landau - Ginsburg - like
model is then invalid for such self-segregating reactive systems
while the CGMD remains a valid MicRep of the system.

This situation is also representative for the typical
biological systems where the same genes
and proteins lead spontaneously to different reaction chains
in different spatial regions of the cell/organism.

Another example is in linguistics in which the carriers
of languages (speakers) are usually spatially
segregated and the transitions between various languages
are quite sharp rather than through a continuum of
interpolating dialects.

Similar effects appear in the ecology of
various species populations.

The coarse grained MicRep rather than the continuum
approach is also necessary if one is interesting to
go beyond continuous deformations of solids
and study the appearance and spreading of cracks
and fractures.

{\bf psychophysics}

In psychology and psychophysics the identification
of the appropriate microscopic elementary objects
is still problematic but it was
shown that it is not beyond the scope our quantitative tools.

As an example we considered the
well defined problem of recognizing
3 dimensional global structure and  motion from the
2 dimensional information available at different times
and locations (on the retina).

It was argued by a combination of rigorous analytical
results supported by a series of experiments \cite{nava}
 that contrary to the previous data,
the 3D structure is re-constructed by our visual system from
point-like signals rather than contour portions passing through the
receptors area.

Moreover, predicting and verifying macroscopic features such as
visual illusions
in certain controlled reproducible conditions,  one was able to validate
 a theoretical model about the way the microscopic signals are
integrated
 by the human visual system into an emerging 3D collective motion
percept.

In particular, it appears that,
in order to lower  the computation time
necessary for producing in real time the 3D percept,
our visual system adopts quite "sloppy algorithms"
which lead to coarse approximations
rely on plain guessing and are often wrong
(e.g. systematically mistake certain rigid motions as non-rigid)
\footnote{
In view of the rigorous limitations
for Turing machines computations on what
can be computed (uncomputability theorems)
and at which price (NP theorems),
this general human brain tendency for sloppines,
resorting to irrational beliefs and
gambling on luck might turn out to be computationally sensible
(and morally commendable).}.

At the present stage, many details are still missing
and the detailed microscopic
representation model for the visual system is far from completed.
However, as mentioned above, the prediction-validation
dialogue has already started.

Some further clues might be provided by the study of
the emergence of contours from the early children
scribbles- a formal and visual analogy
was observed between this process and the protein
folding dynamics \cite{draw}.

The schematic MicRep profile of this application is:
\begin{itemize}
   \item -{\bf Microscopic seers and macroscopic sight} \cite{nava}.
      \begin{itemize}
       \item Micro  - line elements, points.
       \item INTER  - curvature, continuity, integration, (mind) changes
       \item Macro  - shapes (coils, helixes, hashes)
      \end{itemize}
\end{itemize}

We are listing below the MicRep scheme
of a few other  systems proposed in the literature
after which we describe one of them
(the stock market simulation)
in full detail:

{\bf Examples of MicRep's, their Micro's, their  elementary
INTERACTIONS and their Macro's}
\begin{itemize}

   \item -{\bf  microscopic drivers and macroscopic jams} \cite{biham}.
      \begin{itemize}
           \item Micro  - cars
           \item INTER  - go ahead/give way at intersections.
           \item Macro  - traffic flow, jamming.
      \end{itemize}

   \item -{\bf   microscopic Einsteins and the macroscopic universe
        geometry}

     Open Boundary Random Triangulated Surfaces
     and the Wheeler De-Witt Equation for the wave function of the
     2D universe \cite{pazi,esti}.
         \begin{itemize}
         \item Micro  - points, links, triangles
         \item INTER - connectivity of points
         \item Macro  - surface geometry, boundary length distribution
                     (wave function of the universe).
         \end{itemize}

\item - {\bf microscopic Alexander moves and
macroscopic Reasoning/Theorem Proving systems}
       \begin{itemize}
         \item is Simplicial Quantum Gravity Turing Computable \cite{nabu}?
         \begin{itemize}
          \item Micro  - links, letters.
          \item INTER  - simplexes, words.
          \item Macro  - global topology, statements.
         \end{itemize}
        \item  inventive solution search \cite{yanco}.
         \begin{itemize}
          \item Micro  - postulates, statements
          \item INTER  - syllogisms, heuristic rules
          \item Macro  - contradictions, solutions, theorems.
         \end{itemize}
       \end{itemize}
\item -{\bf dramas -
      mathematical categories endowed with dynamics} \cite{gerh}.
         \begin{itemize}
          \item Micro  - categories
          \item INTER  - relations, composition laws
          \item Macro  - (stories) dramas
         \end{itemize}

   \item -{\bf  microscopic investors and macroscopic crashes} \cite{shiki}.
      \begin{itemize}
      \item Micro  - investors, shares
      \item INTER  - sell/buy orders
      \item Macro  - market price (cycles, crushes, booms,
                           stabilization by noise)
      \end{itemize}

\end{itemize}

Let us describe this last MicRep in full detail.

 {\bf economic simulations}

In the study of economic systems,  the simulation
of the money market is usually based on the integration
of a differential stochastic equation in which the
average interest rate and the momentary dispersion
(volatility ) are parameters.
In more involved  estimations of futures values,
various versions of the resulting stochastic dynamics
are considered and an average is performed over the
possible histories.

Another treatment is to construct a network
of simple elements whose interactions do not attempt to
constitute an approximation of the actual microscopic
dynamics of the model. Still, one attempts to obtain
through iterative optimization an "output" of the
system which fits the real macroscopic data.

Yet another technique is to assume that the macroscopic
dynamics can be modeled by a set of a few coupled stochastic
differential equations, to guess their
form and to isolate the universal features in the chaotic
behavior of their solutions.

The MicRep formulation allows a more first-principles
treatment \cite{shiki} by considering explicitly a set of investors which
sell or buy equities according to simple rules of
stochastically maximizing a utility function
which is constructed on the basis of  their previous
history and  prejudice.

The market price of the equity is then computed based
on the bidding prices from all the participating investors.
This in turn influences the wealth of the investors
and their prejudices.

The collective emerging dynamics is by-and-large
independent on the details of  each individual buy-sell
particularities and displays general features.

In particular we found the natural emergence of
cycles of  booms and crashes and the dependence
of their timing and intensity on the
dispersion of the individual buy-sell criteria.

By comparing the standard economic models simulations with the
MicRep one,  one sees that the relation is similar
to the one between thermodynamics and statistical mechanics:
while the thermodynamics can provide a general
macroscopic framework which {\bf relates} the {\bf possible}
values of certain parameters  which
parameterize the macroscopic dynamics, the statistical mechanics
can address more detailed questions and eventually
{\bf deduce } the macroscopic parameters from
the microscopic properties.

{\em The model}

The simple system introduced in \cite{shiki}
contains as its microscopic elements $I$ investors
interacting via buying and selling as detailed below.

The macroscopic dynamics are the trends in the stock market.

Each investor decides at each time how to divide its
wealth $W$ between two investment options:
\begin{itemize}
\item It can invest
in {\bf stock} a proportion  $0 < X(i) < 1$
of its wealth and
\item Invest the rest of it $(1- X(i))$
 in {\bf bonds}.
\end{itemize}
The objective of each investor
(concretized in the strategy detailed below) is to
maximize  the expected utility ${\cal EU} \Bigl ( X(i) \Bigr )$
which in most of our runs we take to be $log W$.

The {\bf bond} is a riskless
investment yielding a constant
return $RR$ after each period.
Thus an investment of $(1- X(i))*W$ dollars will yield
$(1- X(i))*W*(1+RR)$ in the next period.

We actually treat the bonds as depositing money in a bank
at fixed return rate.
Therefore the statements from now on that the bonds are riskless might
be interpreted as dynamically equivalent with requiring
for simplicity the existence
of an ideal riskless (at least very safe compared with the equities)
way of keeping ones money and still getting a return.

This simplifying assumption as well as most of the ones below
can be easily relaxed and their effect can be studied in detail.
In fact the strength of our approach is that
it is very robust to the further introduction of any degree of
realism one deems necessary.

The non-trivial fact is that already in the absence of the
effects of real goods production, prime-matter price fluctuations,
public policies interference, taxation, global politics, etc.
the system presents realistic behavior of its nontrivial
cycles dynamics.

This supports the view that the stock market has a structured
(not pure noise) dynamics by itself, autonomous to a large
extent from the details of its environment.
This is an important prerequisite for any attempt to
predict and characterize stock market future tendencies.

The {\bf stock} is the risky asset.
The return yielded by the stock is composed of two
elements:
\begin{itemize}
\item $P_n$ - the price per share of the stock during
the time period $n$.
$P_n$ is determined collectively by all
investors by the law of supply and demand
as detailed below.
\item The company gives dividends according to its profits.
For simplicity, we assume here that the firm
pays a constant dividend $D$ per share every period
(the effects of variations in $D$ are presently under study).
\end{itemize}
The overall return per share per period is then:
$$H_{n}= {{P_n - P_{n-1} + D} \over P_{n-1} } $$

The investors
prejudices are expressed
by keeping track of the last $K$ returns by the stock.
This we
call the stock's history:
$${\cal H}_n = \{ H_{n}, H_{n-1} ... H_{n-K+1} \}  .$$
For ${\cal H}_0 $ the $H_j$'s with $j = 0, -1, ..., -9$, are given as
part of the initial conditions.
In the initial
model we assumed that the investor believes each one of the
last $K$ history elements $H_j$ has an equal
probability of $1/K$ to occur in the next
period. Later we altered the assumption of equal probability
without observing a qualitative change.

Let us explain in detail the rules which decide
the bid of each investor and the establishing of the
new stock price.

Suppose investor $i$ holds
at time $n$ a number of shares
$N_n(i)$ at the price
of $P_n$ a share,
and has a total wealth (including its
bonds) of $W_n(i)$.
In particular the values
$N_0(i)$, $P_0$ and respectively $W_0(i)$
are provided as initial conditions.

Before each trade period,
each investor $i$ decides
according the procedure described below
how many shares $N_{n+1}(i)$ he wishes to hold
as a function of the hypothetical new stock price
$P_h$
which the offer and demand may fix during that
trade period (according to the procedure
detailed later).
Its bid is formalized in a function
$N_h (i, P_h)$ which expresses how many
shares he orders for each hypothetical value
of the market price $P_h$.
It has to specify a function rather than a firm
shares number, because according to the bidding rules,
the actual value of the price $P_{n+1}$
resulting from the trade round is not yet known
at the time of its bid.

The procedure followed by each investor to compute
$N_h (i, P_h)$ as a function of $P_h$ is
based on its attempt to maximize
the utility function at time $n+2$.
Since $N_n$ is already fixed,
it is too late at this stage to try to affect the wealth
at time $n+1$. It is going to be
$$W_h(i) = W_n (i) + N_n(i)(P_h - P_0)  $$
independently on the $N_{n+1}$ value.
Therefore, in general, before the $(n+1)^{-th}$ trade period,
one tries to optimize the utility function:
$$ U_{n+2} \Bigl ( X_{n+1}(i)\Bigr ) =
\log \biggl [\Bigl(1-X_{n+1}(i)\Bigr)W_h(i)(1+RR) +
{X_{n+1}(i)W_h(i)} \Bigl (1 + H_{n+1}\Bigr )\biggr ] $$

The first term in the $U_{n+2}$ formula
is the contribution of the bond
while the second term is the stock's contribution.
Of course the catch is that the investor does not know
the value of
$H_{n+1}$ ($H_1 $ in the first round case) in this term.
To supply for it, one maximizes in place of
 $ { U}_{n+2} \Bigl ( X(i)\Bigr )$ a mean value of
${ U} \Bigl ( X(i)\Bigr )$ 's
obtained by substituting respectively $H_{n+1}$
with each of the $H_j$'s belonging to ${\cal H}_n$ into the
$ U_{n+2} \Bigl ( X(i)\Bigr )$ formula:
$$ {\cal EU}_{n+2} \Bigl ( X(i)\Bigr ) =
1/K\sum_{j=0}^{-K+1} \log \biggl [\Bigl(1-X(i)\Bigr)W_h(i)(1+RR) +
{X(i)W_h(i)} \Bigl (1 + H_{n+j}\Bigr )\biggr ] $$
The optimum proportion of
investment $X(i) = X_{n+1}(i) $
depends therefore of the knowledge and use
which the investor makes of the history
${\cal H}_n$ and of how it compares with $RR$.

The optimum proportion  of investment $X_{n+1}(i) $
determines the number of shares:
since $ X_{n+1}(i)W_h(i) $ is the amount of money investor $i$
wishes to holds in shares at price $ P_h $,
the number of shares is:
$$ N_h(i, P_h) = {X_{n+1}(i)W_h(i) \over P_h} $$
The number of shares investor $i$ wishes to hold as function
of the price of the share is its personal  demand
curve.
Summing the demands of all investors gives us the collective
demand curve.
Since the number of shares in the market, denoted by
$N$, is fixed, the collective demand curve  sets the new price
of the share $P_{n+1}$ by solving
$$ N= \sum_{i=1}^I N_h (i, P_h)  $$
with respect to $P_h$.

After the trading the wealth of each investor has changed:
$$ W_{n+1}(i) = W_n(i) + N_n(i)(P_{n+1} -P_n) $$
Its new wealth equals its old wealth plus the change in price multiplied
by the number of shares he had {\bf before } the trade. The investors
buys and sells shares at market value and does not gain or lose from this.
It does gain or lose money on the shares he {\bf holds }.
Therefore the decision on $X_n (i)$ affects only the wealth at $W_{n+1}$.

The number of shares held by investor $i$ after the trade is:
$$ N_{n+1}(i) = {X_{n+1}(i)W_{n+1} \over P_{n+1}} =
 {X_{n+1}(i)\biggl( W_n(i) + N_n(i)\Bigl(P_{n+1} -P_n\Bigr)\biggr) \over
 P_{n+1}} $$

Now follows a period of no trade at the end of which dividends
and interest are received.
During this period investor $i$ holds $N_{n+1}(i)$
shares, and has
 $$\biggl( W_n(i) + N_n(i)(P_{n+1} -P_n) \biggr )\biggl (1-X_{n+1}(i) \biggr)
$$
invested in bonds.

After this period its wealth is :
$$W_{n+1}(i) = W_n(i) + N_n(i)\Bigl(P_{n+1} -P_n\Bigr) + N_{n+1}(i)D $$
$$+ \biggl( W_n(i) + N_n(i)(P_{n+1} -P_n)\biggr)\biggl (1-X_{n+1}(i)
\biggr)RR$$
where the third term is due to the dividends,
and the fourth due the interest on
the bonds.

The new price $P_{n+1}$ also adds a new element to the history
of the stock.
The most recent return will be:
$$H_{n+1} = {{P_{n+1} - P_n + D} \over P_n } $$
We can now update the history ${\cal H}_n$ to
${\cal H}_{n+1}$ by adding $H_{n+1}$
to the list and discarding $H_{n-K+1}$.

 The model described so far is deterministic.
The 'decision
making' is done by maximizing the expected utility.
We take into account the variability in the behavior
of the various investors by  adding
a random variable to the optimal proportion of investment.
For instance, we replace $X(i) $ with
$$ {\cal X}(i) = X(i) + \epsilon(i) $$
where $ \epsilon(i) $ is drawn at random from a normal distribution
with standard deviation  T .
$ \epsilon(i) $ is drawn separately for each investor.
This randomness plays the part of the 'noise' or
'temperature' in the system.
'High temperature' means a large  T  which means a wider
distribution and a larger deviation from the deterministic  case.

{\em Computer Experiments and Results}

In the simulations described here \cite{shiki}
we chose the period of time to
be one year.
Accordingly we chose the rest of the parameters as follows.

We took the yearly riskless return $ RR=0.1 $.

We chose the elements of ${\cal H}_0$ randomly with mean $ 0.1001 $
and variance $ 0.024 $, so that
the proportion of investment in the first round will be around $ 50\% $.
We took the history (or rather- memory) length $K=10$.
The number of investors was $I = 100 $
(small for computational reasons, but large enough
to exhibit macroscopic phenomena).
The total number of shares was $N = 10,000 $.

The initial wealth of each investor was $W_0 (i) = \$ 1,000 $
(this is arbitrary as long as the initial price
of the share is chosen accordingly).
Initial price of share was  $P_0 = \$ 4.40 $
(chosen so that the first return by the stock
will be close to the mean of the returns in ${\cal H}_0$).
We choose the dividend $D = \$ 0.3 $
so that $D / P_0  = \$ 0.068  $
i.e. a little less than RR.

As mentioned before, there was no fine tuning of the parameters.

Our results are shown in graphs 7 - 11.
To understand these results let us first examine
the deterministic case $ ( T = 0 ) $ in graph 7 .
We see that the price of the stock
rises sharply and then grows at a steady exponential rate.
Let us first concentrate
at the sharp rise. The first rate of return from the stock is relatively high.
This produces a distribution ${\cal H}_1$ which is
higher than ${\cal H}_0$.
This means that investors are willing to increase
there proportion of investment in the equity.
This increase in the proportions of investment (demand),
especially near the maximum proportion
of investment, cause an immediate  increase in the stock price.

After this sharp rise, the stock's history of returns
${\cal H}_n$ is very promising, and the proportions of
investment in equities are fixed at their maximum.
It can be shown, that under the condition
of fixed investment proportions and zero temperature (no noise) :
$$ H_n = {{P_n - P_{n-1} + D} \over P_{n-1} }
= RR +{D \over P_n} \Bigl ( {1
 \over { 1-X}}
\Bigr) $$
We see that as $ P_n  \rightarrow  \infty      ,$   $ H   \rightarrow RR $.
This explains the constant slope of the exponential climb because :
$$ {P_n \over P_{n-1}} \approx {{P_n - P_{n-1} + D} \over P_{n-1} } \rightarrow
 RR $$
when $P_{n-1} $ is large.

We also learn from this result that the differences between the
$H$'s within the history ${\cal H}$
decrease as the average  value of the $H$'s approaches
$RR$ - the fixed interest rate.
Because $H$ reaches $RR$ from above in the limit
$ P \rightarrow \infty $,
the stock remains slightly preferable to the bond all the way.
This situation, however, is highly unstable
against the slightest noise.
The instability is easy to understand:
because the history of returns is
only very slightly preferable to the bond,
the smallest
fluctuation in price can turn the tables and make the bond preferable.
Because the variance
in histories is very small
( all the returns are close to RR ) a small change in
 one of the
 histories causes a dramatic change in the proportions of investment.
As the
proportions of investment drop the price drops sharply
producing a catastrophic crash.

In the total absence of noise ($T=0$) the small fluctuation
which could trigger the crash never happens.
However, for even very small $T$
such a fluctuation will eventually happen and will
be immediately amplified into a very spectacular crash.
It is enough
that one investor should deviate  from the deterministic behavior,
and decide
to lower its proportion of investment in stocks a little,
to  cause the small
fluctuation that will trigger the crash.
This is exactly what happens when we
'turn on the heat', see graph 8.
One can see that the exponential climb of the
 stock
preceding the crash is not as smooth as before.
The small temperature gives rise
to small fluctuations in price.
One of these fluctuations is large enough and
causes the crash. The typical time for the crash
to take place depends on the temperature and
diverges when $T \rightarrow 0$.

After the first crash
the price of the stock reaches rock bottom.
When the price becomes very low,
the fixed dividend ensures that the returns
on the stock will be relatively high.
After a few periods there are enough high
returns in the history to make the stock an attractive investment again,
the
proportions of investment grow and the cycle begins all over again.

Note here a phenomenon reminding us  of the recent history
of certain authoritarian economic regimes:
by enforcing artificially a narrow volatility
(by strict control of the market) one could postpone
indefinitely the crash (the time scale being related
as we saw above to the narrowness of the dispersion
function).
However, the longer the crash is postponed,
the deeper and more catastrophic it will be.

In contrast, a noisy "free market" displays
a "stabilization by noise" effect
i.e. it is quite jumpy  on short range
but it has rather mild long range instabilities.
Moreover we have found
the stability is helped
if the periods of exchange and the
quantization of the minimal transaction
(share) are denser.

Indeed, in graph 9  one can see what happens when
we turn on the heat.
The temperature adds the random element to the deterministic  run
and has the effect of "smearing" the cycles,
but underneath the noise one can
still see the cyclic behavior.
As the temperature grows the sharp rises and crashes become smoother.
At first glance the prices in Fig. 10 ($T=1.2$) seem random,
but a closer look reveals the traces of cycles,
in fact they
resemble patterns from the real stock market (Fig. 11).

Our MicRep confirms the phenomenological observation
that the ratio of the  dividend yield to price
is a reliable indicator for state of the market.
In particular we confirm that
when the dividend yield is relatively low it is a sign of a bear
market and a crash is to be expected.
The opposite holds if the dividend yield is high,
exactly as we have found in our simulations.

Another common belief
which finds a numerical echo in
our experiments  is that program trading is to be blamed for
the 1987 crash.
When many investors follow the same investment
strategy a strong crash is more probable.
Indeed we obtained strong
crashes when we assumed homogenous decisions.
However,
once we 'turned on the heat' and created heterogeneous decision making
the cycles became milder and the crashes are much smaller.
The higher the
temperature the lower the probability for a dramatic crash.

The implication for the optimal market structure
is that the larger the number of brokers and investment consultants
in the market and the the diversity of their operation,
the smaller the chance of a crash.
If there is one a guru in the stock market
sharp fluctuations in the market are expected.

 It is encouraging to discover such a
wealth of phenomena rising from such a simple
model.
It is even more encouraging that these phenomena seem to fit reality,
and other, completely different, macroscopic models so nicely.
The model can be enriched by introducing different stocks,
investors with different utility functions,
borrowing, bankruptcy, fluctuating dividends and so on \cite{shiki}.

One can study also the breaking of symmetry which takes place when
one starts from a state with all investors equally wealthy
and with identical criteria. Still, any
initial small random fluctuations may get amplified.
Does this increased polarization in wealth lead to increasing
market instability?

As in most of the MicRep applications we mentioned in this review,
the ability to model and simulate the system
based on its most basic elements
brings about deeper yet more direct and intuitive  understanding
of the macroscopic behavior we observe.

\section{  Visualization and Animation of MicRep as tools to identify
the CSD modes and the Relevant Macros}

As mentioned in the introduction, we regard the
elimination of CSD as the numerical
analog to an exact solution
in the analytical treatment.

According to this analogy,
the construction of an RCSD algorithm
is an operational proof of the understanding of the
relevant multi-scale mechanisms.

Usually a mathematical proof is only
a consecration of an understanding which precedes and
conditions it.
In the same way, before the construction of a RCSD algorithm,
and quite independent on its very existence
one can accumulate a quite detailed understanding
of the relevant Macros and their dynamics.

Often, analytical proofs might
be intuitively less illuminating than an heuristic explanation
of the considerations which lead to the proof.
In the same way, a series of interactive animation experiments
may prove as valuable as the construction of an RCSD algorithm.

The understanding might be expressed, validated and improved through
an active dialogue between the researcher and the system.
In such a dialogue the researcher makes predictions
about the  effects of various changes which he implements in the system
and
the simulation answers with confirmations of these predictions
or suggests corrections.

This interactive computer simulation-visualization-animation
paradigm was applied lately
to the multi-scale study of various systems in
Quantum Gravity, Random Triangulated Surfaces,
Solid-on-Solid systems, Fermions dynamics, Gauge Theories,
Topological Objects,
Molecular Dynamics, Nonlinear Dynamics, Fluid Dynamics,
Psychophysics, Neural Network Learning, Chaotic systems, Drawing
Dynamics, Image Dynamics and a multitude of
Quantum Field Theories and Statistical Mechanics systems
\cite{computercenter,viscre}.

Such scientific experiments,
documented in a series of computer animation movies
\cite{computercenter}
and follow-up papers,  proved the
usefulness of the interactive animation method
in solving multi-scale MicRep problems by:
\begin{itemize}
\item suggesting new (including analytical) formulations,
\item identifying visually the relevant Macros and their dynamics,
\item generating more powerful mental representations of the
Macros including their multi-scale hierarchy.
\end{itemize}

We give below 4 examples.

1) {\it Dynamics of Fermions in Lattice Gauge Background}

The Parallel transported Multigrid (PTMG) method is one of the promising
numerical methods to eliminate CSD in the inversion of the
Dirac operator in Lattice Gauge background \cite{longPTMG}.

The PTMG iterative solver represents the Dirac equation
on a hierarchy of grids with various lattice spacing scales.
During a PTMG iteration, the error, residual and
the solution of the equation
are transported between various grids using the
gauge background field as a parallel transporter
This is analog with the
DAMG procedure in chap. 3 where the fields were transported
between the grids with the help of the $J_{ij}$'s.

However, in PTMG, the gauge fields are also used as parallel transporters
for the gauge background itself in the process of
representing the Dirac equation on the coarse lattice.

The identification of the
CSD modes as the Atyiah-Singer zero-modes
and their relation to large scale spatial features
(corresponding to instantons)
was established in the Parallel Transported Multigrid
method with the initial help of a detailed visual search
by computer animation described below \cite{computercenter,visu}.

The visual study was prompted initially by
the puzzling explosion (rather than convergence) of the
error during the PTMG iterations in certain gauge backgrounds.

We suspected that the problem might be with the
well known difficulties of representing faithfully
instantons and the associated fermionic Atyiah-Singer zero-modes on
the lattice.

The visualization studies proved the contrary:
both the gauge and the fermion fields of the continuum
are quite
faithfully represented by the lattice (modulo the doubling problem)
and the
representations on lattices with various lattice spacings
scales are quite consistent one with another.

We identified visually the zero modes
related to the continuum Atyiah-Singer theorem
and discovered that, in the PTMG runs which diverge,
the diverging "error" is always proportional
to those zero-modes.
Moreover we verified that PTMG transports faithfully
the geometrical aspect of the
error, residual and solution between fine and coarse grids
and vice-versa.

It turned out that
the divergence which alarmed us in the first place,
rather than indicating a deficiency
of the PTMG method, was the result of the presence of some
zero-modes which are $exactly$ represented on certain
lattices. These modes lead to a degenerate family of solutions
of the Dirac equation on those lattices.
This family is then spanned by the "diverging" runs.

Based on these facts, and on a theoretical analysis
of the chiral symmetry on the lattice,
we predicted that at $m=0$, for instantons of dimensions
$(4 n) \times (4 n)$) the PTMG process is strongly divergent,
for instantons of dimensions
$(2n+1)\times (2n+1)$ the process is strongly convergent,
while for
$(4n+2)\times (4n+2)$ the error approaches very quickly
a constant and gets totally stuck there.

These predictions were dramatically confirmed by our subsequent runs
\cite{visu}.

As a by-product of this visualization study, we were left with
a significantly increased understanding of the geometrical
properties of the lattice Dirac equation and of the
Dirac equation on compact spaces in general.

2) {\it The Wave-Function of the  2D Quantum Universe}

This visualization was prompted by the study
of the open-boundary RTS model
which is a model for the 2 dimensional quantum gravity
\cite{triang}.

In order to simulate the RTS model , one has to
generate stochastically triangulated surfaces composed of $N$ triangles
such that each triangulation with local 2-dimensional
topology has equal chance to be generated.
This distribution is realized usually by generating
the triangulations recursively using a local algorithm
(Alexander move) which updates the position of
one link at a time.

We wished to measure the probability distribution $P(L)$ of
the length $L$ of the boundary.
P(L) corresponds in the continuum limit to
the wave function of the 2D universe.
The objective was to compare it with certain analytical
predictions from the continuum
\cite{eliezer,nati}.

We realized that the measured \cite{pazi} mean value of the length
$$< L > \equiv \int L P(L) d L \approx 0.765 N$$
and the width of its distribution
$$ \sigma (L) \equiv \int (L- <L>)^2 P(L) d L \approx 0.65 \sqrt { N }$$
suggest a geometry in which most of the triangles
touch the boundary and
there is a random density of "branching" triangles
(triangles where 3 otherwise disjoint
surface regions are joint together).

By visualizing the triangulated "surfaces",
we observed that this $P(L)$ distribution
is the result of the
fact that the typical open "surface" looks rather like a
tree of beads of width $1$ which are branching at every scale.
The branches are CSD Macros under the usual
Alexander moves dynamics.
This is due to the fact that
the large scale branching features are topologically metastable
under local geometry changes.

Using the intuitions emerging from this picture,
we were able to write a rigorous recursion relation \cite{esti} which
allowed us to estimate certain amplitudes
(i.e. the relative number of RTS's with a given topology)
which were not available by any other method.

In particular we were able to estimate the amplitude
corresponding to the
"figure eight" diagram \cite{nati}: the amplitude for creating
from the vacuum a pair of baby universes which touch in 1 point
(the ancestor of a foam).

It turned out that this diagram represents the
(long searched-for) solution
to the Wheeler-DeWitt equation
for the wave function of a universe starting from nothing
\cite{esti}.

3) {\it Visual study of the Solid-On-Solid (SOS) roughening simulations}

The SOS are models for the roughening transition of the interface
between 2 phases (e.g. cristal vs. gas, or spin up vs. spin down
regions in the 3D Ising model).
The SOS models are defined on (usually) regular 2 dimensional
lattices. The variables are integers defined on the lattice sites
and represent the "height" (or "depth") of the interface with
respect to a certain reference "horizontal" plane.

The energy of each configuration is minimal when the heights
of neighboring sites are equal and increases with their difference.

The Valleys-to-Mountains-Reflection (VMR) \cite{evertz1} algorithm
views each configuration of the critical SOS system
as a landscape with valleys and mountains as its Macros.
The VMR algorithm creates stochastically new configurations
by imagining that a mountain is sectioned by a horizontal plane
along one of its equal-height contours.
The new configuration is obtained by reflecting about the horizontal
plane the mountain region situated above the plane.
In this way, this region becomes a valley.

A similar transformation is devised for reflecting "valley"
bottoms into "mountains".
Care is taken that the actual procedure respects scrupulously
the detailed balance (\ref{detbal}).

The VMR algorithms \cite{evertz1}
allowed an improvement by an order of magnitude in the precision of
the simulations of Solid-On-Solid (SOS) models and their
Kosterlitz-Thouless transition \cite{elsewhere}.

In order to understand and generalize the successes of VMR
in other contexts,
we produced an animation movie which compares
the dynamics of various VMR versions.
We compared the VMR version which reflects surface regions
with respect to horizontal planes situated at half integers heights
({\bf H}-algorithm)
with the VMR version which reflects with respect to integers
(combined with local Metropolis or HB) ({\bf I}-algorithm).

This allowed us to observe and study the metastable Macros
which are responsible for the remaining CSD in each case.

In the case of the {\bf I} algorithm, the CSD Macros
turned out to be large terraces of height 1.
The reflection about an integer, can
transform such a terrace into a shallow
"pond" of depth 1 with the same shape and position.
However the reflection about an integer
cannot generate, annihilate or deform
such a height 1 / depth 1 feature.

The incapability of the {\bf I} reflections to
change significantly the shape and/or the position of these Macros
is responsible for the CSD and
leads to a dynamical critical exponent
$z\approx 1.$ for the {\bf I} algorithm.

In the case of the {\bf H} algorithm,
the CSD ($z\approx 1.5$) is due to the
incapability to perform efficiently the above
"terrace-pond" switch. However, the {\bf H}
reflection is capable of generating and annihilating
step 1 features.

The combined algorithm
(alternating {\bf I} and {\bf H} steps)
includes all the relevant
Macro moves and consequently it eliminates completely CSD
($z\approx 0.$).

4) {\it Visualization of the Instanton tunneling dynamics}

One of the simplest lattice gauge theories presenting
topological sectors indexed by an instanton charge $Q$ and
the related Atyiah-Singer fermionic zero-modes is the
2-dimensional U(1) gauge model.

It can be shown that the typical local $Q$ density is about
$1 / (\xi \times \xi)$ where $\xi$ is the correlation length.
Consequently, for a realistic simulation ($\xi \approx  L/8$),
the typical values for the total topological charge
are between $-10$ and $10$.

In the process of studying the fermion propagators
in gauge background, we realized that the usual local
Metropolis or Heat Bath (HB)\cite{binder} algorithms
(defined according tothe rules in chap 3)
never tunnel between $Q$-sectors for realistic
$\xi$ ($ > 5$) values.
In fact, this is consistent with the
theoretical estimation of the lattice
energy barrier which separates the topological sectors \cite{fuc}.

Therefore, we were witnessing an extreme case of CSD where the
Macros were topological instantons.

In order to prepare a representative ensemble of gauge configurations
we had to devise a global algorithm which explicitly
transformed a configuration into another configuration
belonging to a different topological sector \cite{fuc}
while scrupulously fulfilling the detailed balance condition
(\ref{detbal}).

The resulting "instanton-offering" algorithm
produced this time the correct $Q$ distribution consistent
with the analytical estimates \cite{longPTMG}.

In order to study the way the new algorithm
works, we visualized the evolution of the gauge
configurations, and compared the dynamics of the local (HB)
algorithm with the "instanton-offering" algorithm (plus local).

However, the differences between the 2 procedures are not apparent
when the elementary plaquettes values are displayed.
At short distances,
the configurations obtained by the 2 algorithms are undistinguishable
(all one sees is the thermal noise of non topological excitations).

In order to see the effects of the instanton dynamics
one has to display rather, the contributions to $Q$
comming from coarser lattice subsets.
As one displays
 plaquette averages over various scale sub-lattices,
the dynamical effects of the global steps become apparent.
For subsets of the order of $\xi$, the
effects of the tunneling take the form of sudden jumps.
Such jumps, or any other changes,
are never observed in the local algorithm runs
even for sublattices much larger than $\xi$.
In particular the total topological charge never changes.
after the initial thermalization.

We also discovered visually the disturbing fact that,
even though none of the measurements
of the usual quantities shows CSD
after introducing the global instanton-generating steps,
certain metastabilities still persist in the
long range geometrical aspect of the configurations.
These effects might be related to CSD in n-point
correlation functions ($n>2$) and deserve further study.
This is an interesting new situation in which the Macros
are discovered before the CSD they produce.

The 4 examples above are only a sample of the multi-scale
computer visualization experiments we performed for physical and
other systems.

The interactive multiscale animation techniques proved
so powerful that one had sometimes the feeling
that they tap into a short-cut leading directly
to the inner mental representation forms manipulated by
the human mind in the process of understanding.
A possible explanation for it is that the natural
structures inherent to the human thought are intrinsically
multi-scale. As such, they might be considered as
a legitimate target for further MicRep study.

{\bf Acknowledgements}

I would like to thank all the colleagues which
contributed to the shaping of my thinking on the
fascinating subject of this review and especially
R. Ben-Av, A. Brandt, M. Creutz, G. Mack, A. Schwimmer.

I thank also the participants to the
course I gave at SISSA in the spring of 1994
especially D. Amati, L. Bonora and R. Iengo.
I learned more from their questions, than they learned from
my answers.

The present article was completed during my stay with the
Theory Group at the Swansea University where I enjoyed the
hospitality of I. Halliday and D. Olive.

The research leading to this paper was supported in part
by grants from the Germany-Israel Foundation, the Israeli
Academy and by the Welsh Science Foundation.

\newpage
\begin{center}
{\bf \Large Figure Captions}
\end{center}

{\bf Figure 1}:

The decorrelation time $\tau_{\bf{k}} $ versus $k$, in Metropolis
dynamics.
The squares denote
the data points with their corresponding error bars,
and the solid line denotes the best fitting.
Note that different modes ({\bf k}'s) can have the same $k$.\\
The linear lattice size is L=64, $z=2.1\pm 0.05$.

{ \ }

{ \ }

{ \ }

{ \ }

{ \ }

{\bf Figure 2}:

Same as Fig. 1 for L=128,
$z=2.1\pm 0.1$.

{ \ }

{ \ }

{ \ }

{ \ }

{ \ }

{\bf Figure 3}:

The decorrelation time $\tau_{\bf{k}}$ versus $k$, in Wolff
dynamics. \\(a) L=64, $z=0.1\pm0.2$. (b) L=128, $z=0\pm0.15$.

{ \ }

{ \ }

{ \ }

{ \ }

{ \ }

{\bf Figure  4}:

The evolution of $O=<(x-<x>)^2>$
during a DAMG run (crosses) vs.
an usual HB run (continuous line).
One can see that during the DAMG run the $O$ measurements effectively
cover the support of its  relevant distribution while
the HB run does not even reproduce the correct mean value.
The horizontal axis of each run is normalized such
that the same point on the axis corresponds to equal CPU times.
The lattice size was $8596$ triangles.

{ \ }

{ \ }

{ \ }

{ \ }

{ \ }

{\bf  Figure 5}:

The logarithm of the time auto-correlation function $C$,
for the observable
$<(x-<x>)^2>$ is plotted as a function of the number of sweeps $n$.
The results from DAMG runs (crosses) and HB runs (diamonds)
appear on the same graph for comparison.
The ordinates $n$ of the two graphs
are respectively normalized such that
each point on the horizontal axis represents equal CPU times
for DAMG and HB.

{ \ }

{ \ }

{ \ }

{ \ }

{ \ }

{\bf Figure 6}:

The logarithm of the time auto-correlation function $C$,
for the observable
$<(x-<x>)^2>$ is plotted as a function of the number of sweeps $n$.
The results from DAMG runs (crosses) and HB runs (diamonds)
appear on the same graph for comparison.
The ordinates $n$ of the two graphs
are respectively normalized such that
each point on the horizontal axis represents equal CPU times

{ \ }

{ \ }

{ \ }

{ \ }

{ \ }

{\bf Figure 7}:

 The the evolution of the stock value in
the "deterministic" case $ ( T = 0 ) $.

{ \ }

{ \ }

{ \ }

{ \ }

{ \ }

{\bf Figure 8}:

The stock cycles in the presence of a small
random dispersion of the market.

{ \ }

{ \ }

{ \ }

{ \ }

{ \ }

{\bf Figure 9}:

The stock cycles in the presence of a larger
random dispersion of the market.

{ \ }

{ \ }

{ \ }

{ \ }

{ \ }

{\bf Figure 10}:

The stock cycles for a realistic random dispersion
($T=1.2$) of the market.

{ \ }

{ \ }

{ \ }

{ \ }

{ \ }

{\bf Figure 11}:

The real stock market evolution during the last decades.
\end{document}